# ONLINE ADAPTIVE FAULT TOLERANT BASED FEEDBACK CONTROL SCHEDULING ALGORITHM FOR MULTIPROCESSOR EMBEDDED SYSTEMS

Oumair Naseer[1] and Rana Atif Ali Khan[2]

[1]Department of Computer Science, University of Warwick, Coventry, United Kingdom
o.naseer@warwick.ac.uk
[2]School of Engineering, University of Warwick, Coventry, United Kingdom
Atif.khan@warwick.ac.uk

*ABSTRACT*

*Since some years ago, use of Feedback Control Scheduling Algorithm (FCSA) in the control scheduling co-design of multiprocessor embedded system has increased. FCSA provides Quality of Service (QoS) in terms of overall system performance and resource allocation in open and unpredictable environment. FCSA uses quality control feedback loop to keep CPU utilization under desired unitization bound by avoiding overloading and deadline miss ratio. Integrated Fault tolerance (FT) based FCSA design methodology guarantees that the Safety Critical (SC) tasks will meet their deadlines in the presence of faults. However, current FCSA design model does not provide the optimal solution with dynamic load fluctuation. This paper presented a novel methodology of designing an online adaptive fault tolerant based feedback control algorithm for multiprocessor embedded systems. This procedure is important for control scheduling co-design for multiprocessor embedded systems.*

## KEYWORDS

*Feedback based control scheduling algorithm, Multiprocessor Embedded Systems, Fault Tolerance and Control Scheduling Co-design,*

## 1. INTRODUCTION

Use of control theory in real time embedded systems design has increased massively over the past few years, and this trend keeps on evolving day by day [1]. Due to the large number of real time constrains and requirements, the design complexity of feedback based control co-design of multiprocessor embedded systems has increased and over 90% of the embedded controllers are used to control real time processes and deceives[2]. Scheduling is the key lever in real time computing system for overall system performance and resource utilization. Traditional scheduling algorithms used in embedded system design are Rate Monotonic (RM) and Early Deadline First (EDF). From the control point of view, all these classic scheduling algorithms are open loop [10] and these algorithms are designed based on the assumption that mapping of the jobs/tasks is predefined and Worst Case Execution Time (WCET) of jobs is known a priori. Due to the open and uncertain environment, the overall execution time of both safety critical and non safety critical tasks varies. It is very difficult to predict actual timing constraints of the task before execution. To avoid this uncertainty, feedback based control scheduling algorithms are employed in control system co-design of real time multiprocessor embedded systems [11] [12, 13, and 14]. FCSA combines the feedback based control theory in hardware/software co-design of embedded systems, so that the available resources can be used optimally and the overall performance of the system can be increased.

Faults associated to multiprocessor embedded systems can occur either in hardware or in software. These faults are categorised into (i) transient faults: occur only for a short period of time and (ii) permanent faults: affects the system everlastingly [3, 4]. Traditional Fault tolerant schemes are based on the hardware redundancy [2 and 5] and can avoid only a single transient or a single permanent fault. This method incurs high hardware cost to add a new functionality. On the other hand, FT schemes can be implemented in software as well. Most promising FT schemes are; (i) Active replication, in which a task is replicated on two or processors and replicas perform the required services [6]. (ii) Re-execution; in re-execution whenever a fault is detected, task is re-executed from the start which increases execution overhead to a large extent. (iii) Primary back up; in this scheme each task has a backup when a fault is detected, backup task is executed to perform the required services. (iv) Check pointing [7]; in check pointing Safety Critical task is divided into *n* sub-tasks and each sub-task contains a check point appended by either a programmer [8] or by the compiler [9]. Fault is detected based on these check points. In case of fault, there are two options either to roll back or roll forward. This scheme is helpful in avoiding the transient faults. From the scheduling point of view, a combination of active replication and re-execution provides more optimized system design and better CPU performance.

## 2. RELATED WORK

For soft real time computing systems, a feedback performance control is presented in [16] which primarily focus on applying control theory to real time scheduling and utilization control. A state of the art feedback control scheduling algorithm for real time computing systems with variable execution time is presented in [17] which provide the performance guarantee for hard real time tasks. Feedback based Dynamic Voltage Scaling (FDVS) method to select proper frequency and voltage for Fault tolerant hard real time embedded system is presented in [32, 35]. Author also tries to provide QoS by reducing energy consumption and satisfying hard real time constraints in the presence of transient faults. It also provides a technique to integrate DVS with control theory for hard real time embedded systems. An analysis of distributed feedback control with shared communication and resources utilization for real time computing system is addressed in [19]. Integrated Fault tolerance scheme check-pointing for real time embedded systems is presented in [7]. A perspective on integrating feedback control and computing for control scheduling co-design is addressed in [18].Feedback control design for networked control system; a novel approach for designing feedback based control scheduling for the networked systems is presented in [20, 21]. Up to date feedback control scheduling algorithms based on Fuzzy logic controller for network control is presented in [12]. An adaptive neural network based feedback control scheduling for real time computing systems is presented in [13 and 14]. In [11], author presented an approach to recover system from fault mode for parallel systems using check-pointing FT scheme. A Trade offs between fault tolerant schemes and control theoretical method is presented in [33, 34]. In [15], author provides a double feedback based control scheduling approach for real time computing systems to optimize overall system performance. A feedback based control scheduling for hard real time systems is addressed in [18] but this work doesn't address the online adaptation. Feedback based control scheduling co-design approach for real time embedded systems is presented in [20] and this work shows that closed loop systems are not hard real time systems. Although, control systems are more robust in nature and uncertain to time variations but they also suffers from time jitters and data loss. In [22, 23], author tires to capture the time variation of Safety Critical (SC) tasks over network for better resource utilization in correspondence with sampling intervals and time delays to achieve QoS in terms of CPU performance. System response in presence of Fault and recovery schemes for hard real time systems to achieve dependability in X-by-Wire (XBW) systems is addressed in [29 and 30].

A fault tolerant scheduling for hard real time embedded system is addressed in [31], but this work only focuses on maintaining CPU scheduling with specified scheduling bound by making sure that SC tasks will meet their deadlines. Moreover, this work doesn't capture the state of the task in Fault mode and provides less information about data loss. To the best of our knowledge, this is the first work that addresses online adaptive feedback based control scheduling and fault tolerance together for multiprocessor embedded systems.

## 3. PROBLEM STATEMENT

The primary objective of integrated FT based FCSA is to provide QoS in terms of CPU performance and resource utilization, by keeping CPU utilization at schedulable bound in the presence of faults. The design methodologies of integrated Feedback based scheduling algorithms are based on the separation of the concerns [15]. These concerns are derived from the assumptions that feedback controllers can be designed by assuming the fixed predefined mapping, hard deadlines and fixed time period. These assumptions are widely used in the control community because they help the control embedded system designer to design control loops without concerning the nature of the overall system in the presence of faults. This paper presented a new methodology of designing an online adaptive fault tolerant based feedback control algorithm for multiprocessor embedded systems which provides better CPU performance and resource utilization.

## 4. MULTI PROCESSOR SYSTEM ARCHITECTURE

System architecture constitutes a distributed shared Hardware (HW) platform with a network topology [24, 25], where every hardware node can communicate with every other HW node. Fig. 1 shows the high level multiprocessor system architecture model and resources elaborating the partitioning concepts. It also describes the application execution environment, where HW nodes are connected through a network bus. Each HW node has two cores; one core is completely dedicated for SC tasks and second one is dedicated for the non SC tasks [27-29]. Each node has a capability of executing both safety critical and non safety critical tasks. Node resource consists of an I/O controller, CPU, sensors and actuators, RAM, ROM and a Feedback based scheduling Controller (FSC). Every HW node in integrated multiprocessor system architecture utilizes the same configuration. Feedback based control scheduling algorithm is implemented on the top of RTOS layer. It is assumed that the allocations of tasks are predefined and faults can occur at any time.

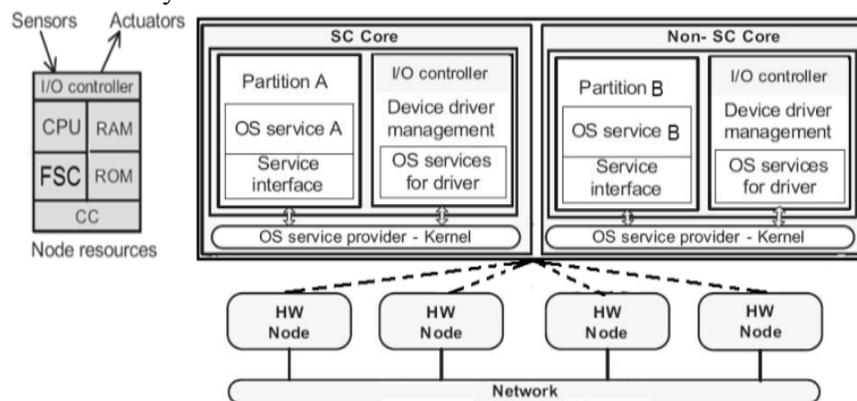

Figure 1: Integrated system architecture: Tasks of mix-criticality (SC and non SC) executes on the same node.

## 5. FEEDBACK BASED CONTROL SCHEDULING MODEL

FCSA is implemented as a set of tasks running on top of an off-the-shelf Real Time Operating System (RTOS) using fixed-priority and pre-emption. Control performance in terms of stability and tracking error relies on the values of sampling rates and sensors to actuators latencies. From the control theory point of view, multiprocessor embedded systems are non-linear in nature and are usually modelled by a set of periodic tasks assigned to one or several processors [26]. A Worst Case Execution Time (WCET) technique is used to analyse fixed-priority real-time computing systems. Task periods are the main actuators of the control system running on the top of a fixed priority scheduler with the aim to adjust on-line sampling periods of the controllers in order to meet the computing resource requirements and CPU utilization. Control inputs variables are the periods of the control tasks and output variable is the measured CPU utilization as shown in the below Fig 2.

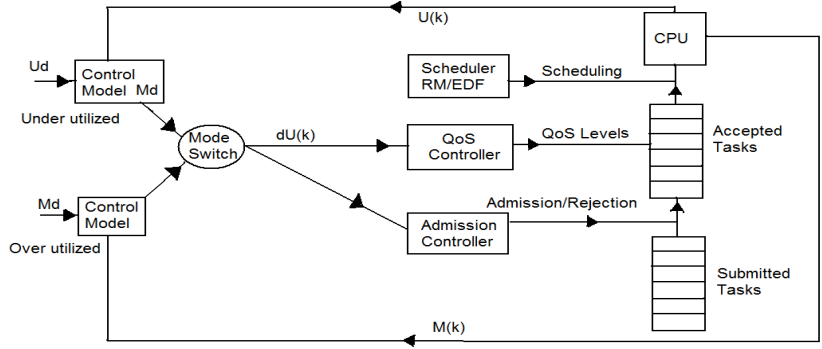

Figure 2: Feedback Control Scheduling Architecture.

Controller model design is flexible and well known approaches are Proportional Integral Differential (PID) controller, Linear Quadratic (LQ) controller, Fuzzy logic controller and adaptive neural network. U(k) is the total CPU load measured for each period of scheduling task and M(k) is the task deadline miss ratio. Ud is the desired load and Md is the controller variable, to control the task deadline miss ratio. Adding Feed-forward admission controller allows future tasks cost anticipation and for enhanced transient behaviour.

Processor utilization model is defined in the following equation which holds for any number of processors [22, 31].

$$y(t+1) = Ay(t) + B\Delta r(t) \qquad (1)$$

Where $y \in L^n$ represents the processor utilization vector with size $n$; $\Delta r \in L^m$ represents the change to task execution rate from the $m$ number of tasks running on the processor. $B \in L^{n \times m}$, and is defined as;

$$B = D\,K \qquad (2)$$

Where $K$ is the available subtask allocation matrix that record which number of particular tasks are running on which processors. $D = diag\{d_1, d_2, \ldots, d_n\}$ is a diagonal matrix, and $d_i$, *where i=1,2,3...n,* are scalar values that denote the ratio between the change to the actual utilization of processor $i$ and its estimation $\Delta r(t)$. The size of $d_i$ measures the estimation error, i.e., how much the actual execution time of each task on processor $i$ deviates from its estimated value.

## 6. ONLINE ADAPTIVE CONTROLLER DESIGN

Online adaptive control mainly consists of a Linear Quadratic (LQ) controller and a Recursive Least Square model estimator (RLS) as shown in the below figure.

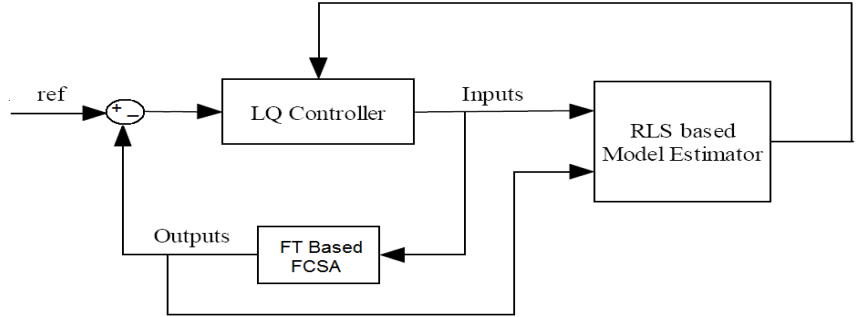

Figure 3: Online adaptive controller model.

RLS based Model Estimator learns and update the LQ model of the FT based FCSA. A reference point is set to keep the output at the desired value. This is done by setting the control inputs by minimizing a quadratic cost function.

### 6.1. RLS-based Model Estimator

FT based FCSA is a multiprocessor embedded system and can be modelled as a multiple-input-multiple-output (MIMO) as follows:

$$L(q^{-1})y(k) = M(q^{-1})u(k) + d(k) \qquad (3)$$

Where $L(q^{-1})$ and $M(q^{-1})$ are matrix polynomials in the back-ward shift operators.

$$L(q^{-1}) = l - L_1 q^{-1} - \ldots - L_n q^{-1} \qquad (4)$$

$$M(q^{-1}) = M_0 q^{-1} - M q^{-1} - \ldots - M_{n-1} q^{-1} \qquad (5)$$

where $l$ is the order of the FT based FCSA multiprocessor system, $d(k)$ is a sequence of independent, identically distributed *n-dimensional* random vectors with zero mean representing disturbances. We assume that $d(k)$ is independent of $y(k-s)$ and $u(k-s)$ for $s > 0$.

$u(k) = \Delta r(k)$ is the control input which is the vector of estimation task execution rate change, and $y(k)$ is the control output which is the vector of processor utilizations. RLS based model estimator with exponential forgetting estimates the coefficient matrices $i\ L$ and $s\ M$ online, where $0 < i < 1$ and $0 \leq s < l$, and their values keep on changing due to varying runtime conditions. CUP utilization model equation can be re-written as;

$$y(t+1) = X(t)z(t) + d(t+1) \qquad (6)$$

Where

$$z(t) = [u^T(t) \ldots u^T(t-l+1) \quad y^T(t) \ldots y^T(t-l+1)]^T \quad and$$
$$X(t) = [M_0, \ldots M_{i-1}, L_1, \ldots L_l] \qquad (7)$$

RLS estimator with exponential forgetting identifies the time varying parameters of matrix $X(t)$ and is defined as:

$$B(t+1) = y(t+1) + \hat{X}(k)z(t) \qquad (8)$$

Where

$$\hat{X}(t+1) = X(t) + \frac{B(t+1)z^T(t)P(t-1)}{\lambda + z^T(t)P(t-1)z(t)} \quad \text{and}$$

$$P^{-1}(t) = P^{-1}(t-1) + \left(1 + (\lambda-1)\frac{z^T(t)P(t-1)z(t)}{(z^T(t)z(t))^2}\right)z(t)z^T(t) \quad (9)$$

where $\hat{X}(t)$ is the estimation of the $X(t)$; $B(t)$ is the estimation error vector, $P(t)$ is the covariance matrix; $\lambda$ is the forgetting factor $0 < \lambda < 1$.

## 6.2. Linear Quadratic (LQ) Optimal Controller

The primary objective of online adaptive controller is to let the FT based FCSA output track the reference command with small tracking error. by avoiding large changes to the control inputs. This is done by minimizing the quadratic cost function $A$ defined as follows:

$$A = \left\|V\left(y(t+1) - y_{ref}(t+1)\right)\right\|^2 \quad (10)$$
$$+ \|Q(u(t) - u(t-1))\|^2$$

where $V$ is a positive-semi-definite weighting matrix on the tracking errors, (a higher weight indicates higher importance value of the corresponding output variable). $Q$ is a positive-definite weighting matrix to penalize large changes in the control inputs. $V$ and $Q$ are defined as diagonal matrices and their relative magnitude provides a way to trade-off tracking accuracy for smaller changes in the control input.

## 7. EXPERIMENTS AND RESULTS

The purpose of the first experiment is to keep the CPU utilization at the desired set point = 0.8123 without knowledge of actual task execution time. For this experiment, g=0.30 for both processors which means actual execution time is 30% of the estimated time. Also initial task rates are assigned based on the estimated execution times to make the utilization equal to the set point.

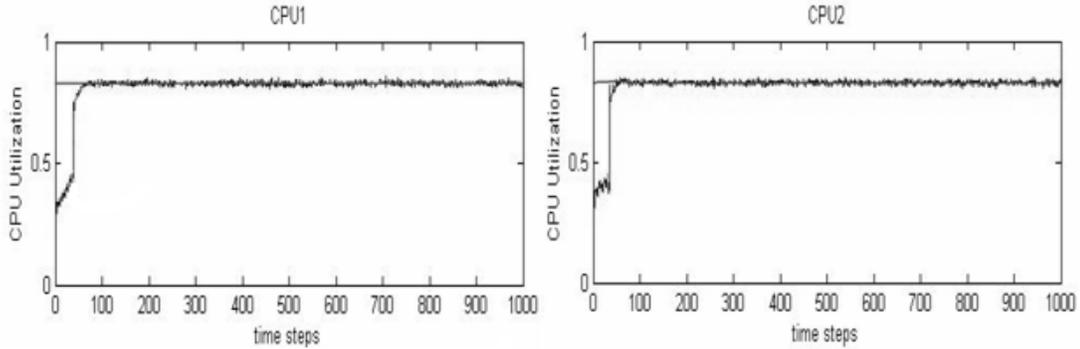

Figure 4: Result of experiment 1: CPU utilization at desired set point 0.8123.

Fig. 4 shows the processor utilization responses. Both processors are underutilized initially. The task rates are then increased gradually until the utilization of both processors converges to the set point of 0.8123.

The purpose of the second experiment is to find the upper bound on the estimated execution time. For this g is set to 7 for both CPUs which means that the actual execution time is seven times the estimated value. Both processors are initially over-utilized. The task rates are then decreased gradually until the utilization of both processors converges to the set point of 0.8123.

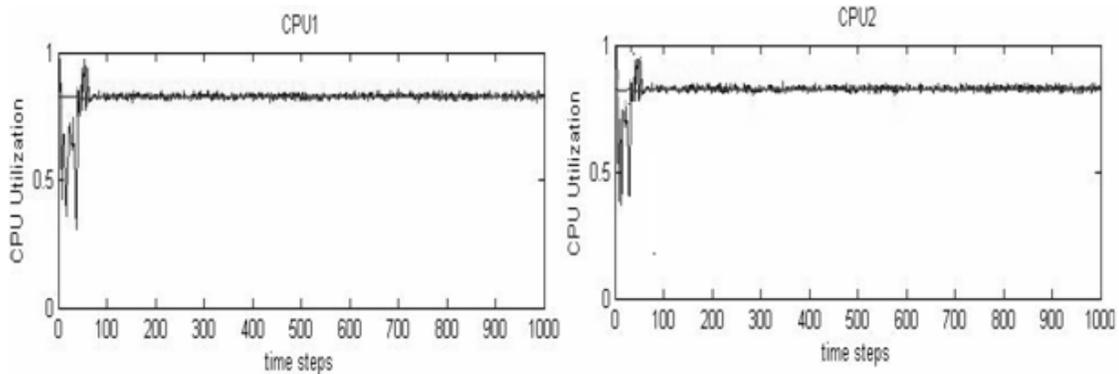
Figure 5: Result of experiment 2: CPU Over utilized condition.

Fig. 5 shows the processor utilization responses. CPU utilization exhibits some initial oscillations due to model estimation inaccuracies, but as model estimation becomes more accurate later and they converge to the utilization set point 0.8123 quickly.

The purpose of third experiment is to investigate the robustness of online adaptive controller, for this the task execution rate is varied dynamically and the CPU utilization is again set to 0.8123.

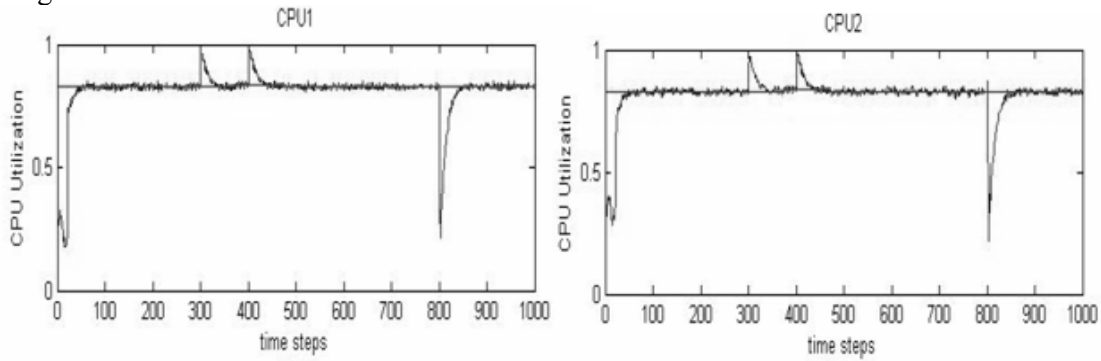
Figure 6: Result of experiment 3, investigating robustness (load fluctuation) of controller.

Fig. 6 shows the processor utilization responses. When the workload is changed at the $300^{th}$, $400^{th}$ and $800^{th}$ sample steps, the online adaptive controller keeps the utilizations at the desired set point 0.8123 with very smaller oscillation.

## 8. CONCLUSION AND FUTURE CONSIDERATIONS

In this paper, an online adaptive Fault tolerance based Feedback control scheduling algorithm for multiple embedded systems is presented. The CPU model is investigated form different perspective by first keeping the CPU under-utilized and then increasing the task rate initially, then keeping CPU over-utilized and then decreasing task rate gradually. Finally, the robustness of the system is investigated by dynamically varying the task execution time. The overall system model is more stable and provides Quality of Services in terms of CPU performance and resource usage. For g=7, which means the actual execution time is seven times the estimated value, the system remains stable with little oscillation and all tasks meet their deadlines. However, if the task execution time increases beyond this value, the system no longer remain stable and oscillation results in the deadline miss of tasks.

Feedback scheduling has become an important methodology in dynamic co-design of control and scheduling for real time multiprocessor embedded systems. With different structures and algorithms, it enables better use of the computing resources and leads to better CPU performance. In future, a more improved CPU utilization model and advance hybrid online controller may result in better overall system performance and resource usage. However, the practical implementation of feedback scheduling-based control systems is an almost completely open issue.

## 9. REFERENCES

[1] B. Bouyssounouse, J. Sifakis, Embedded Systems Design: The ARTIST Roadmap for Research and Development, Springer, 2005.

[2] P. Agrawal. Fault tolerance in multiprocessor systems without dedicated redundancy, IEEE transactions on computers, 37:358-362, March 1988,

[3] P. A. Bernstein. Sequoia: A fault-tolerant tightly coupled multiprocessor for transaction processing, Computer, 21:37-45, February 1988.

[4] J-C., Laprie, & B. Randell, Basic Concepts and Taxonomy of Dependable and Secure Computing, IEEE Transactions on DependableSecure Computing (TDSC), 1(1), pages 11{33, 2004.

[5] R. M. Keichafer, C.J. Walter, A.M. Finn & P.M. Thambidurai, The MAFT Architecture for Distributed Fault Tolerance, IEEE Transactions on Computers, 37(4), pages 398{405, 1988.

[6] S. Poledna, P. Barrett, A. Burns, & A. Wellings, Replica Determinism and Flexible Scheduling in Hard Real-Time Dependable Systems, IEEE Transactions on Computers, 49(2), pages 100{111, 2000.

[7] S. Poledna, P. Barrett, A. Burns, & A. Wellings, Replica Determinism and Flexible Scheduling in Hard Real-Time Dependable Systems, IEEE Transactions on Computers, 49(2), pages 100{111, 2000.

[8] Avi Ziv, jehoshua Bruck, Analysis of checkpointing schemes for multiprocessor systems, 13th Symposium on Reliable Distributed Systems, 1994.

[9] K. M. Chandy and C. V. Ramamoorthy, Rollback and recovery strategies for computer programs, IEEE Transactions on computers, 21:546-556, June 1972,

[10] J. Long, W. K. Fuchs, and J. A. Abraham. Fowrawd recovery using checkpointing in parallel systems. In the 19th International Conference on Parallel Processing, pages 272-275, August 1990.

[11] C. Lu, J.A. Stankovic, G. Tao, S.H. Son, "Feedback control real-time scheduling: framework, modeling, and algorithms", Real-time Systems, Vol.23, No.1/2, pp. 85-126, 2002.

[12] Sha, L., T. Abdelzaher, K.-E. Årzén, T. Baker, A. Burns, G. Buttazzo, M. Caccamo, A. Cervin, J. Lehoczky, A. Mok, "Real-time scheduling theory: A historical perspective", Real-time Systems, Vol.28, 2004.

[13] A. Goel, Walpole, and M. Shor. "Real-rate scheduling," in proceedings of the 10th IEEE Real-Time and Embedded technology and Applications Symposium (RTAS), pp. 434-441, 2004.

[14] S. Lin and G. Manimaran. "Double-Loop Feedback-Based scheduling Approach for Distributed Real-Time Systems," in proceedings of the High Performance Computing (HiPC), pp. 268-278, 2003.

[15] J.A. Stankovic, T. He, T.F. Abdelzaher, M. Marley, G. Tao, S.H. Son, and C. Lu. "Feedback Control Real-TimeScheduling in Distributed Real-Time Systems," in proceedings of the IEEE Real-Time Systems, 2001.

[16] K.E. Årzén, B. Bernhardsson, J. Eker, A. Cervin, K. Nilsson, P. Persson, and L. Sha, Integrated control and scheduling. Technical Report ISRN LUTFD2/TFRT7586SE. Lund Institute of Technology, Sweden, 1999.

[17] C.L. Liu and J.W. Layland, "Scheduling Algorithms for Multiprogramming in a Hard Real-Time Environment," J. ACM, vol 20,no. 1, pp. 46-61, 1973.

[18] C. Lu, J.A. Stankovic, G. Tao, and S.H. Son, "Feedback Control Real-Time Scheduling: Framework, Modeling, and Algorithms,"Real-Time Systems J., vol. 23, no. 1/2, pp. 85-126, 2002.

[19] Feng Xia and Youxian Sun, Control-scheduling codesign: A prespective on integrating control and computing. Dynamics of Continuous, Discrete and Impulsive Systems - Series B, vol. 13, no. S1. 2008

[20] Jianguo Yao and Xue Liu, Mingxuan Yuan, Zonghua Gu, Online Adaptive Utilization Control for Real-Time Embedded Multiprocessor Systems, ACM, 2008.

[21] Payam Naghshtabrizi and João P. Hespanha. Analysis of Distributed Control Systems with Shared Communication and Computation Resources, American Control Conference, 2009.

[22] J. Liu, Real-Time Systems: Prentice Hall PTR 2000.

[23] C. Lu, X. Wang, and K. X., "Feedback utilization control in distributed real-time systems with end-to-end tasks," Parallel and Distributed Systems, IEEE Transactions on, vol. 16, no. 6, pp. 550-561, 2005.

[24] CAN Specification, Controller Area Network Specification and Implementation, Robert Bosch GmbH, http://www.semiconductors.bosch.de/pdf/can2spec.pdf, 1991.

[25] The FlexRay Group, FlexRay Communications System Protocol Specification, Version 2.1, http://www.°exray.com/, 2005.

[26] Daniel Simon, NeCS-INRIA and Alexandre Seuret NeCS-CNRS Peter Hokayem and John Lygeros, Eduardo Camacho, State of the art in control/computing co-design. The Joint Laboratory for Petascale Computing (JLPC). 2010.

[27] C. Wilwert, N. Navet, Y.-Q. Song & F. Simonot-Lion, Design of Automotive X-by-Wire Systems, In The Industrial Communication Technology Handbook, CRC Press, 2004.

[28] V. Claesson, S. Poledna & J. Soderberg, The XBW Model for Dependable Real-Time Systems, International Conference on Parallel and Distributed Systems (ICPADS), pages 130{138, 1998.

[29] X-by-Wire Project, Brite-EuRam 111 Program, X-By-Wire – Safety Related Fault Tolerant Systems in Vehicles, Final Report, 1998.

[30] J. P. Hespanha, P. Naghshtabrizi, and Y. Xu, "Survey of recent results in networked control systems," Proc. of IEEE, vol. 95, no. 1, pp. 138–62, Jan. 2007.

[31] P. Naghshtabrizi, "Delay impulsive systems: A framework for modeling networked control systems," Ph.D. dissertation, University of California at Santa Barbara, Sep. 2007.

[32] A. Jhumka, M. Hiller, & N. Suri, Assessing Inter-Modular Error Propagation in Distributed Software, IEEE Symposium on Reliable Distributed Systems (SRDS), pages 152{161, 2001.

[33] Ali Sharif Ahmadian, Mahdieh Hosseingholi, and Alireza Ejlali, A Control-Theoretic Energy Management for Fault-Tolerant Hard Real-Time Systems, Real-Time Systems Symposium (RTSS), 2011.

[34] S. Ghosh, R. Melhem, and D. Mosse, "Fault-Tolerant Scheduling on a Hard Real-Time Multiprocessor System," in Proc. 8th Int. Symp. Parallel Processing, pp. 775-782, 1994.

[35] Y. Zhang and K. Chakrabarty, "Dynamic Adaptation for Fault Tolerance and Power Management in Embedded Real-Time Systems," ACM Trans. Embedded Computing Systems, vol. 3, no. 2, pp. 336-360, 2004.